\documentclass[aps,pra,reprint,groupedaddress,showkeys]{revtex4-1}
\usepackage{amsmath,graphicx,hyperref,verbatimbox}
\usepackage[FIGTOPCAP,raggedright,nooneline,bf,footnotesize]{subfigure}
\usepackage{indentfirst}
\usepackage{braket}
\usepackage{diagbox}
\usepackage{amssymb}
\usepackage{bbold}
\usepackage{colortbl}
\usepackage{multirow}
\newtheorem{thm}{Theorem}[section]
\newcommand{\figref}[1]{Fig.~\ref{#1}}
\newcommand{\tabref}[1]{Tab.~\ref{#1}}
\renewcommand{\eqref}[1]{Eq.~\ref{#1}}
\newcommand{\tr}{{\mathrm{Tr}}}

\begin{document}

\title{Quantum Tomography of Pure States with Projective Measurements Distorted by Experimental Noise}
\author{Artur Czerwinski}
\email{aczerwin@umk.pl}
\affiliation{Institute of Physics, Faculty of Physics, Astronomy and Informatics \\ Nicolaus Copernicus University,
Grudziadzka 5, 87--100 Torun, Poland}

\begin{abstract}
The article undertakes the problem of pure state estimation from projective measurements based on photon counting. Two generic frames for qubit tomography are considered -- one composed of the elements of the SIC-POVM and the other defined by the vectors from the mutually unbiased bases (MUBs). Both frames are combined with the method of least squares in order to reconstruct a sample of input qubits with imperfect measurements. The accuracy of each frame is quantified by the average fidelity and purity. The efficiency of the frames is compared and discussed. The method can be generalized to higher-dimensional states and transferred to other fields where the problem of complex vectors reconstruction appears.
\end{abstract}
\keywords{quantum state tomography, mutually unbiased bases, complex vector reconstruction, phase retrieval}
\maketitle

\section{Introduction}

Quantum state tomography (QST) originated in 1852, when G.~G.~Stokes derived the polarization state of a light beam based on intensity measurements \cite{Stokes1852}. The problem of state identification remains relevant since well-characterized quantum resources are required for quantum information processing \cite{Kus2001} and quantum key distribution (QKD) \cite{Beige2002}. Apparently, there are many approaches to QST which differ from one another in the kind of measurement(s) and the number of their repetitions \cite{dariano03,paris04}. One popular method involves polarization measurements and can be applied to recovering the quantum state of photons \cite{James2001}. Other frameworks connected with photonic state tomography rely on registering Hong-Ou-Mandel interference \cite{Wasilewski2007,Kolenderski2009}.

A fundamental problem of QST has always concerned determining a set of measurement operators sufficient to identify an unknown state \cite{Dariano01}. Some theoretical proposals aim at reducing the number of operators to the necessary minimum \cite{Czerwinski2016,Czerwinski2020}, while experimental frameworks tend to apply overcomplete sets of operators in order to overcome noise and errors \cite{Horn2013,Horn2019}.

To clarify the problem investigated in the article, let us postulate that the achievable information about a $d-$level pure quantum system is encoded in a complex vector -- called the state vector and denoted by $\ket{\psi}$, which belongs to the Hilbert space $\mathcal{H} \cong \mathbb{C}^d$ such that $\dim \mathcal{H} = d < \infty$. Moreover, $\ket{\psi}$ is normalized, i.e. $\braket{ \psi | \psi} = 1$, where $\braket{ .|.}$ denotes the inner product in the Hilbert space. Then, the goal of QST is to reconstruct the accurate representation of $\ket{\psi}$ on the basis of data accessible from an experiment. Naturally, multiplying the state vector by a scalar of unit modulus does not change the measurement results. Thus, the state vector can be determined up to a global phase factor.

In the case of pure states tomography, we are analyzing the problem of recovering a complex vector from intensity measurements -- the very same kind of problem is considered in many other areas of science, from pure mathematics to speech recognition or signal processing \cite{Jaganathan2016}. Thus, there is vast literature concerning phase retrieval, see a review article Ref.~\cite{casazza14}. In particular, in recent years, a lot of attention has been paid to the connection between complex vector reconstruction and the theory of frames, e.g. Ref.~\cite{balan06,Jamiolkowski2010}.

Let us recall a definition. By an $M-$element complex frame in $\mathbb{C}^d$, denoted $\Xi = \{ \ket{\xi_1}, \dots,\ket{\xi_M} \}$ (where $\ket{\xi_i} \in \mathbb{C}^d$), one should understand a set of complex vectors that span $\mathbb{C}^d$. In articles not connected to quantum tomography, authors usually consider in general the problem of reconstructing an unknown complex vector $\ket{x} \in \mathbb{C}^d$ from its intensity measurements, i.e. it is discussed whether the knowledge about the non-linear map:
\begin{equation}\label{eq:1}
\mathcal{J}_{\Xi}: \ket{x} \rightarrow \left( \left|\braket{\xi_i |x}\right|^2 \right)_{i=1,\dots,M}
\end{equation}
is sufficient to determine the complex vector $\ket{x}$. In Physics, this type of measurement is referred to as projective measurement.

To formulate a sufficient condition for complex vector reconstruction, let us first revise general definitions involving the question when phase retrieval is possible. In Ref.~\cite{conca15}, the authors propose to assume that phase retrieval is possible when any two vectors $\ket{\psi}$ and $\ket{\psi'}$ with identical intensity measurements differ only by a scalar of norm one, i.e. $\ket{\psi} = e^{i \phi'} \ket{\psi'}$. In other words, the same postulate can be stated that it is possible to reconstruct a complex vector $\ket{\psi}$ if and only if the non-linear map $\mathcal{J}_{\Xi}$ is injective and $\Xi$ is a frame. Thus, henceforth, in the situations when the phase retrieval is possible, we shall say that the frame $\Xi$ generates (or defines) injective measurements.

In Ref.~\cite{bandeiraa14}, Bandeiraa, Cahill, Mixton and Nelson postulated a conjecture according to which if one wants to reconstruct a vector $\ket{x} \in \mathbb{C}^d$, then a frame that contains less than $4d-4$ vectors cannot generate injective intensity measurements, i.e. according to the authors fewer than $4d-4$ modulus of inner product of $\ket{x}$ with other vectors is not sufficient to obtain the structure of $\ket{x}$. Furthermore, in the same paper the authors postulated the second part of the conjecture that a generic frame with $4d-4$ vectors (or more) generates injective measurements on $\mathbb{C}^d$. The second part of the conjecture has been proved in \cite{conca15}, where the authors explained the notion of a generic frame and demonstrated that for a generic frame $\Xi$ which contains at least $4d-4$ elements the corresponding map $\mathcal{J}_{\Xi}$ is injective.

Another recent paper \cite{vinzant15} proves a result that contradicts the first part of the conjecture from \cite{bandeiraa14}. C.~Vinzant proposed a frame in $\mathbb{C}^4$ which consists of $11$ vectors and proved that it defines injective measurements on $\mathbb{C}^4$. Therefore, the current knowledge about the phase retrieval problem does not give an answer to the question what is the minimal number of elements of the frame $\Xi$ so that the map $\mathcal{J}_{\Xi}$ is injective, i.e. so far it remains unknown, in general, how many intensity measurements are required to reconstruct an unknown complex vector. However, in Ref.~\cite{bandeiraa14}, the authors proposed a relatively efficient way to verify whether a given frame $\Xi$ generates injective measurements. Their approach is presented below as a theorem.

\begin{thm}[Bandeiraa et al. 2014]
\label{linearspace}
A frame $\Xi = \{\ket{\xi_1}, \dots, \ket{\xi_M}\}$ (where $\ket{\xi_i } \in \mathbb{C}^d$) defines injective measurements, i.e. one can reconstruct an unknown vector $\ket{x}\in \mathbb{C}^d$ from intensity measurements $\left|\braket{ \xi_i | x}\right|^2$ for $i=1,\dots,M$, if and only if the linear space
\begin{equation}\label{eq:13}
\mathcal{L}_{\Xi} := \{ Q\in \mathbb{C}^{d\times d} : \bra{\xi_1} Q\ket{ \xi_1 } = \dots = \bra{ \xi_M} Q \ket{\xi_M} = 0 \}
\end{equation} 
does not contain any non-zero Hermitian matrix of the rank $\leq 2$.
\end{thm}

The Theorem \ref{linearspace} states clearly the necessary and sufficient condition that needs to be satisfied so that the frame $\Xi$ defines injective measurements and, therefore, it is possible to reconstruct a complex vector on the basis of the intensity measurements. For a given frame, one can relatively easy verify whether the condition stated in Theorem \ref{linearspace} is fulfilled or not. However, so far there has been no concrete proposal concerning a feasible procedure to obtain such a sufficient frame.

In this article, we investigate qubit state reconstruction by two frames which differ in the number of elements. One is defined by the elements from the symmetric, informationally complete, positive operator-valued measure (SIC-POVM) \cite{Renes2004} and the other consists of the vectors from the MUBs \cite{Wootters1989}. The accuracy of each frame is quantified by two figures of merit -- the average fidelity and purity, which are computed and presented on graphs. In Sec.~\ref{method}, we present the framework of QST for pure states and assumptions concerning experimental noise. Then, in Sec.~\ref{results}, the results are introduced and discussed. The findings demonstrate how efficient the frames are at overcoming the experimental noise.

\section{State reconstruction framework}\label{method}

In this work, we assume that the initial state of a qubit can be presented as a vector:
\begin{equation}\label{m1}
\ket{\psi_{in}} =\begin{pmatrix}  \cos \frac{\theta}{2} \\ \\ e^{ i \phi} \sin \frac{\theta}{2} \end{pmatrix},
\end{equation}
where $ 0\leq \phi < 2 \pi$ and $0\leq \theta \leq  \pi$. An unknown quantum state of the form \eqref{m1} can be reconstructed from projective measurements, which in the case of photons are based on photon counting. Let us denote a frame by $\Xi = \{ \ket{\xi_1}, \dots\}$, where $\ket{\xi_i} \in \mathbb{C}^2$. Next, we take into account experimental noise connected with each measurement. In particular, we impose the Poisson noise, which is a typical kind of uncertainty arising in photon counting \cite{Hasinoff2014}. Thus, assuming that the total number of photons equals $\mathcal{N}$, we shall introduce a formula for the measured photon count associated with the $k-$th projective measurement:
\begin{equation}\label{m2}
n^{M}_k = \mathcal{N}_k \,\left|\braket{ \xi_k | \psi_{in}}\right|^2,
\end{equation}
where $\mathcal{N}_k$ stands for a number generated randomly from the Poisson distribution characterized by the expected value $\mathcal{N}$.

For a given frame $\Xi$, we can numerically generate experimental data for any specific input state \eqref{m1} (we shall consider a sample consisting of $400$ input states). However, in quantum state reconstruction problem, we postulate that there is no \textit{a priori} knowledge about an unknown state. For this reason, an output state, which results from the tomographic algorithm, is assumed to be represented by a density matrix \cite{James2001,Altepeter2005}:
\begin{equation}\label{m3}
\rho_{out} = \frac{T^{\dagger} T}{\tr\: \{T^{\dagger} T\}}, \hspace{0.35cm}\text{where}\hspace{0.35cm}T=\begin{pmatrix}t_1 & 0 \\ t_3 + i\,t_4 & t_2 \end{pmatrix},
\end{equation}
which is equivalent to the Cholesky factorization. In other words, the problem of state reconstruction of qubits means that we strive to estimate the values of the four parameters $t_1, t_2, t_3, t_4$ which fully characterize the density matrix. Consequently, the expected photon count for the $k-$th frame vector takes the form:
\begin{equation}\label{m4}
n^{E}_k = \mathcal{N} \,\tr \{ \ket{\xi_k}\! \bra{\xi_k} \rho_{out}  \}.
\end{equation}
In order to estimate the parameters $t_1, t_2, t_3, t_4$, we shall apply the method of least squares (LS) \cite{Opatrny1997}. This method has been used to study the performance of QST frameworks with simulated measurement results \cite{Acharya2019,SedziakKacprowicz2020}. According to the LS method, the minimum value of the following function needs to be determined:
\begin{equation}\label{m5}
\begin{aligned}
{}&f_{LS} (t_1, t_2, t_3, t_4) = \sum_k \left( n^{E}_k  - n^{M}_k  \right)^2 = \\& =\sum_k \left(  \mathcal{N} \,\tr \{ \ket{\xi_k}\! \bra{\xi_k} \rho_{out}  \} - \mathcal{N}_k \,\left|\braket{ \xi_k | \psi_{in}}\right|^2 \right)^2,
\end{aligned}
\end{equation}
which allows one to perform QST for any input state $\ket{\psi_{in}}$ and a given frame $\Xi$.

Next, a scenario with extended measurement errors will be considered. Apart from the Poisson noise, dark counts shall be taken into account. In practice, it means that the detector receives not only the desirable signal, but also some number of photons which come from the background. Mathematically speaking, the background noise shall be modeled by adding to the input state a component proportional to the maximally mixed state (perturbation term), i.e.
\begin{equation}\label{m6}
\rho_{in} = (1- \epsilon) \ket{\psi_{in}} \!\bra{\psi_{in}} + \frac{\epsilon}{2} \mathbb{1}_2,
\end{equation}
where $\mathbb{1}_2$ denotes the $2 \times 2$ identity matrix and $\epsilon$ shall be referred to as the noise parameter (naturally: $0\leq \epsilon \leq 1$). This gives a modified formula for the measured photon counts:
\begin{equation}\label{m7}
\begin{aligned}
n^{M'}_k {}& = \mathcal{N}_k \,\tr \{ \ket{\xi_k}\! \bra{\xi_k} \rho_{in}\} = \\& = (1- \epsilon) \mathcal{N}_k \,\left|\braket{ \xi_k | \psi_{in}}\right|^2 + \mathcal{N}_k \frac{\epsilon}{2},
\end{aligned}
\end{equation}
which can be substituted into the function \eqref{m5} in order to estimate the state in the other scenario.

In this article, we shall compare the quality of qubit estimation with two frames: one defined by the SIC-POVM and the other by the MUBs. First, the scenario with the Poisson noise alone will be considered. In the next step, the perturbation term will be added to the input states. In order to evaluate the efficiency of the frames, we utilize the notion of quantum fidelity, denoted by $\mathcal{F}$, and purity, represented by $\gamma$, which are defined as \cite{Nielsen2000}:
\begin{equation}\label{m8}
\begin{split}
&\mathcal{F} := \left(\tr \sqrt{\sqrt{\rho_{out}}\ket{\psi_{in}} \!\bra{\psi_{in}} \sqrt{\rho_{out}}} \right)^2 \\
&\gamma :=  \tr \{ \rho_{out}^2\}.
\end{split}
\end{equation}
The first figure, $\mathcal{F}$, quantifies the overlap between the actual state $\ket{\psi_{in}}$ and the result of estimation $\rho_{out}$ \cite{Jozsa1994}, whereas $\gamma$ measures how close the reconstructed density matrix is to the pure state. In the case of both measurement scenarios, we perform QST with each frame for a sample of $400$ qubits defined as \eqref{m1}, with $\phi$ and $\theta$ covering the full range. Then, the performance of the frames is discussed based on the average fidelity, $\mathcal{F}_{av}$, and purity, $\gamma_{av}$, computed over the sample, cf. Ref.~\cite{SedziakKacprowicz2020}.

\section{Results and discussion}\label{results}

The goal of this section is to compare the performance of two frames in QST of pure states. One can utilize symmetric, informationally complete, positive operator-valued measures (SIC-POVMs) in order to reconstruct an unknown quantum state \cite{Rehacek2004}. For $\dim \mathcal{H} =2$, we assume that $\{\ket{0}, \ket{1}\}$ denotes the standard basis in $\mathcal{H}$. Then, the SIC-POVM consists of four projectors defined by the vectors:
\begin{equation}\label{r1}
\begin{split}
&\ket{\xi^{SIC}_1} = \ket{0}, \hspace{0.5cm} \ket{\xi^{SIC}_2} = \frac{1}{\sqrt{3}} \ket{0} + \sqrt{\frac{2}{3}} \ket{1}, \\
&\ket{\xi^{SIC}_3} = \frac{1}{\sqrt{3}} \ket{0} + \sqrt{\frac{2}{3}} e^{i \frac{2 \pi}{3}} \ket{1},\\
&\ket{\xi^{SIC}_4} = \frac{1}{\sqrt{3}} \ket{0} + \sqrt{\frac{2}{3}}  e^{i \frac{4 \pi}{3}} \ket{1}.
\end{split}
\end{equation}
Mathematically speaking, the vectors \eqref{r1} constitute a frame which defines injective measurements according to Theorem \ref{linearspace}. The frame comprising the SIC-POVM shall be denoted by $\Xi^{SIC}$.

The other frame, denoted by $\Xi^{MUB}$, consists of six vectors which correspond to the elements of the mutually unbiased bases (MUB) for, i.e.:
\begin{equation}\label{r2}
\begin{split}
&\ket{\xi^{MUB}_1} = \begin{pmatrix} 1 \\ 0 \end{pmatrix}, \hspace{0.5cm} \ket{\xi^{MUB}_2} = \begin{pmatrix} 0  \\ 1 \end{pmatrix},\\
&\ket{\xi^{MUB}_3} =\frac{1}{\sqrt{2}} \begin{pmatrix} 1 \\ 1 \end{pmatrix}, \hspace{0.5cm} \ket{\xi^{MUB}_4} = \frac{1}{\sqrt{2}} \begin{pmatrix} 1  \\ -1 \end{pmatrix},\\
&\ket{\xi^{MUB}_5} = \frac{1}{\sqrt{2}} \begin{pmatrix} 1 \\ i \end{pmatrix}, \hspace{0.5cm} \ket{\xi^{MUB}_6} =  \frac{1}{\sqrt{2}} \begin{pmatrix} 1  \\ -i \end{pmatrix}.
\end{split}
\end{equation}
The frame $\Xi^{MUB}$ also generates injective measurements. From the physical point of view, intensity measurements associated with the frame $\Xi^{MUB}$ can be realized on photons through polarization measurements since the vectors are commonly used to represent: vertical/horizontal, diagonal/anti-diagonal, right/left circular polarization states, respectively, e.g. Ref.~\cite{Bayraktar2016}.

\begin{table}[h]
\setlength{\tabcolsep}{5pt} 
\renewcommand{\arraystretch}{1.75}
	\begin{tabular}{|c|c|c|c|c|}
\hline
		\multirow{2}{*}{
				\backslashbox[22 mm]{\color{black}$\mathcal{N}$}{\color{black}frame}} & \multicolumn{2}{c|}{$\Xi^{MUB}$} & \multicolumn{2}{c|}{$\Xi^{SIC}$} \\ \cline{2-5} 
	 	&  $\mathcal{F}_{av} (\mathcal{N}) $ & $\gamma_{av} (\mathcal{N})$ & $\mathcal{F}_{av}(\mathcal{N})$ & $\gamma_{av}(\mathcal{N})$ \\ \hline
	1 &$0.7714$  &$0.9142$ & $0.7313$ &$0.9036$  \\ \hline
5 &$0.9036$  &$0.9251$ & $0.8788$ &$0.9104$  \\ \hline
		10 &$0.9334$  &$0.9354$ & $0.9080$ &$0.9128$  \\ \hline
	25 &$0.9564$  &$0.9406$ & $0.9461$ &$0.9368$  \\ \hline
		50 & $0.9721$ & $0.9597$ & $0.9655$ & $0.9550$ \\ \hline
	100 & $0.9793$ & $0.9679$ & $0.9761$ & $0.9652$ \\ \hline
1\,000 & $0.9940$ & $0.9890$ & $0.9925$ & $0.9864$ \\ \hline
10\,000 & $0.9981$ & $0.9964$ & $0.9979$ & $0.9959$ \\ \hline
	\end{tabular}
	\caption{Average fidelity $\mathcal{F}_{av} (\mathcal{N}) $ and purity $\gamma_{av}  (\mathcal{N})$ in pure state estimation with two distinct frames. The method of least squares was applied with the function \eqref{m5}. Each value was computed as the mean for a sample of $400$ input qubits of the form \eqref{m1}.}
	\label{qubitestimation}
\end{table}

In order to investigate the efficiency of each frame in pure state reconstruction, numerical simulations were conducted, assuming different number of photons involved in measurements. A sample of $400$ input states of the form \eqref{m1} was considered and each state was reconstructed with the measured photon counts distorted by the Poisson noise \eqref{m2}. The results are gathered in \tabref{qubitestimation}.

It was expected that the impact of the Poisson noise should be greater if we utilize fewer photons per measurement. Thus, we can observe that the accuracy of both frames in QST increases along with the number of photons. For $\mathcal{N} = 10\,000$, both frames lead exactly to the unknown state. However, if the number of photons drops down, one can observe a substantial discrepancy between the state obtained from the algorithm $\rho_{out}$ and the original state $\ket{\psi_{in}}$. It is worth noting that, for smaller numbers of photons, $\Xi^{MUB}$ results in better quality of pure state estimation than $\Xi^{SIC}$. This feature is in agreement with the common practice in QST to employ overcomplete sets of measurement operators in order to combat experimental noise. When we increase the number of photons, the figures of merit for both frames converge.

\begin{figure}[h]
	\centering
   \begin{subfigure}
         \centering
         \includegraphics[width=0.95\columnwidth]{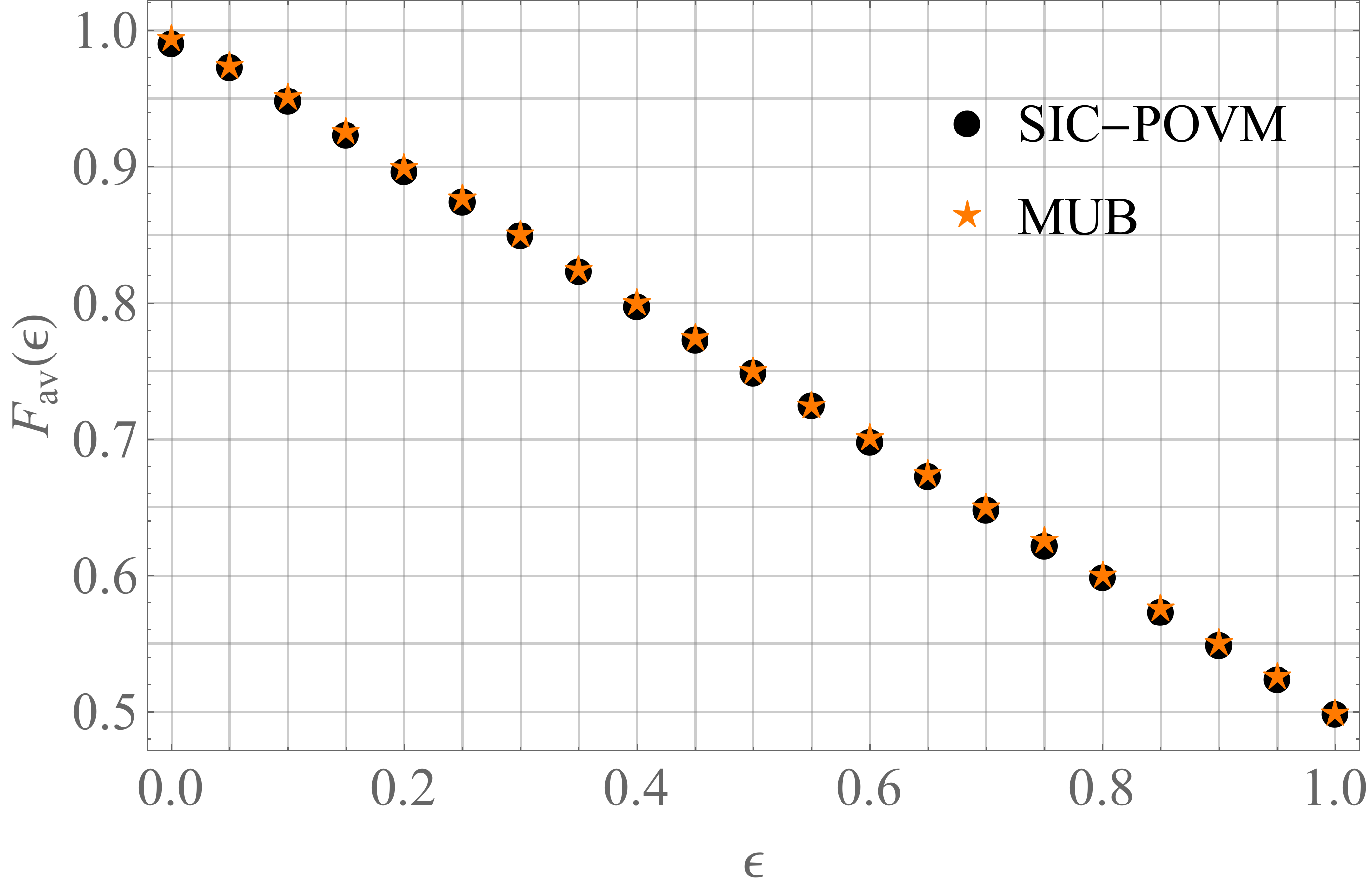}
     \end{subfigure}
     \hfill
     \begin{subfigure}
         \centering
         \includegraphics[width=0.95\columnwidth]{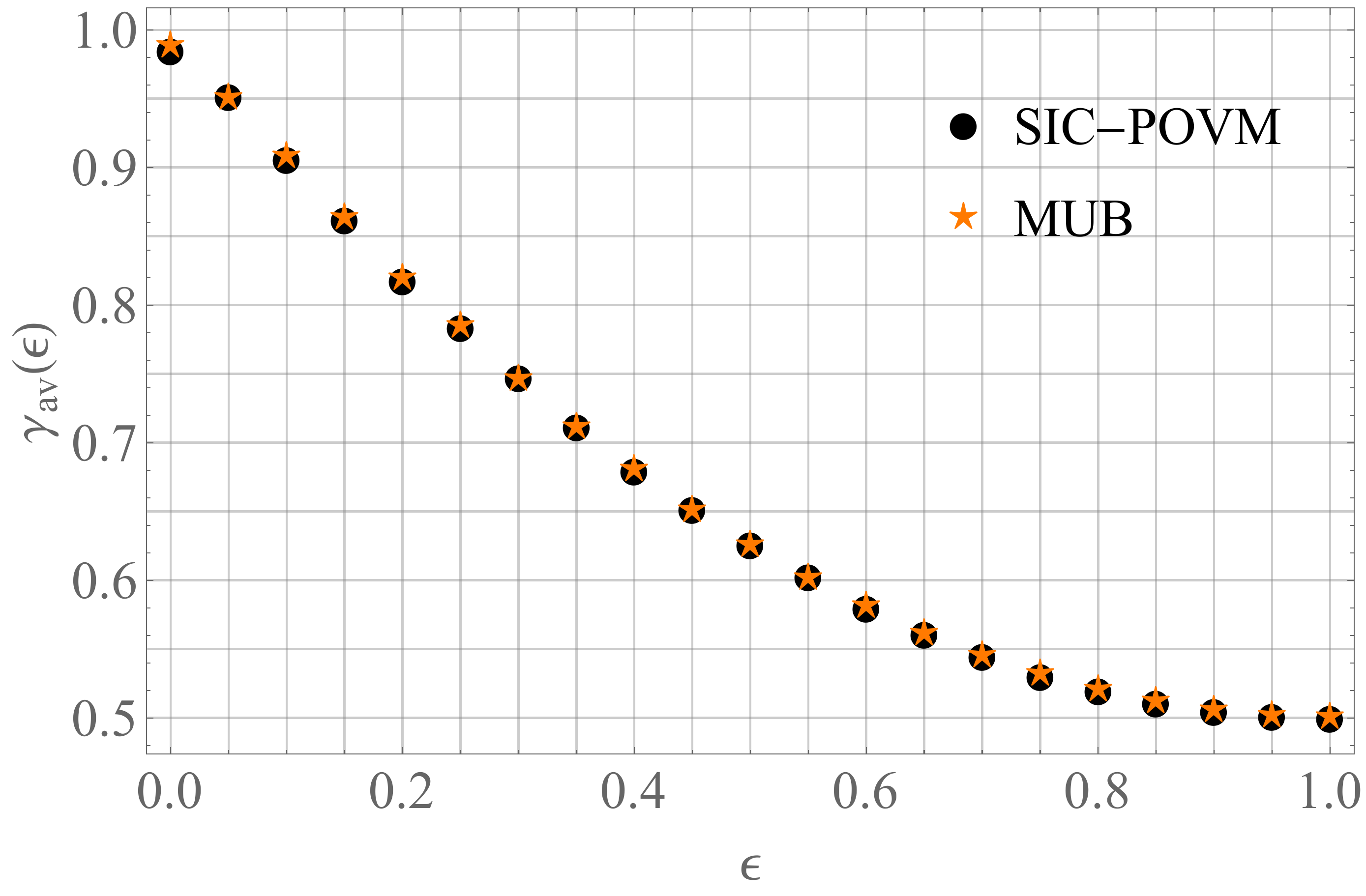}
     \end{subfigure}
	\caption{Plots present the average fidelity $\mathcal{F}_{av} (\epsilon)$ (the upper graph) and the purity $\gamma_{av} (\epsilon)$ (the lower graph). Each point was obtained by the method of least squares for a sample of $400$ input states including a noise parameter $\epsilon$ \eqref{m6}. The formula \eqref{m7} for the measured photon counts was applied, assuming that $\mathcal{N} = 1\,000$.}
	\label{plots}
\end{figure}

Further insight into the efficiency of the frames can be provided by investigating input states influenced by the error parameter $\epsilon$, as in \eqref{m6}. Let us assume that the number of photons is fixed: $\mathcal{N} = 1\,000$. Then, for each frame, we can consider the average fidelity and purity as functions of $\epsilon$, denoted by $\mathcal{F}_{av} (\epsilon)$ and $\gamma_{av} (\epsilon)$, respectively. The plots of the functions are presented in \figref{plots}.

For the number of photons $\mathcal{N} = 1\,000$, the results demonstrate that if we consider the input states distorted by the perturbation term \eqref{m6} (while still keeping the Poisson noise), then both frames lead to very similar accuracy. The whole range of the noise parameter $\epsilon$ was considered and there is no significant difference between the efficiency of the frames. One can notice that the function $\mathcal{F}_{av} (\epsilon)$ is linear, whereas $\gamma_{av} (\epsilon)$ is convex. The plots allow one to observe how the quality of state estimation degenerates as we increase the amount of noise (dark counts).

\begin{table}[h]
\setlength{\tabcolsep}{5pt} 
\renewcommand{\arraystretch}{1.75}
		\begin{tabular}{|c|c|c|c|c|c|}
\hline
		\backslashbox[18 mm]{\color{black}frame}{\color{black}\large{$\epsilon$}} & $0.1$ & $0.2$ & $0.3$ & $0.4$ & $0.5$ \\
			\hline
		$\Xi^{MUB}$ &$0.8960$ & $0.8632$ & $0.8231$ & $0.7867$ & $0.7476$\\ \cline{2-6} 
		 \hline
		$\Xi^{SIC}$ & $0.8919$ & $0.8584$ & $0.8298$ & $0.7900$ & $0.7457$ \\ \cline{1-6}
		\end{tabular}
		\caption{Average fidelity $\mathcal{F}_{av}  (\epsilon)$ in pure state estimation with two distinct frames. Each value was computed as the mean for a sample of $400$ input qubits of the form \eqref{m6}. The formula \eqref{m7} for the measured photon counts was applied with $\mathcal{N} = 10$.}
		\label{fewphotons}
	\end{table}

Finally, let us investigate the difference between the frames in the case of few photons and the presence of dark counts \eqref{m6}. If we decrease the number of photons down to $\mathcal{N}=10$, then from \tabref{fewphotons} we observe that $\Xi^{MUB}$ has no significant advantage over $\Xi^{SIC}$ for non-zero values of the error parameter. Though there are tiny differences between the figures in \tabref{fewphotons}, they should be considered negligible. These results prove that for noisy measurements, distorted by both the Poisson noise and dark counts, $\Xi^{SIC}$ and $\Xi^{MUB}$ deliver the same quality.

\section{Summary and outlook}

In this article, two frames have been compared in terms of their applicability in quantum tomography of qubits. One frame, $\Xi^{SIC}$, comprised the elements of the SIC-POVM and the other, $\Xi^{MUB}$, contained the vectors from mutually unbiased bases (MUBs). Based on the numerical simulations, we have demonstrated that $\Xi^{MUB}$ outperforms $\Xi^{SIC}$ only if we consider a measurement scenario which involves single-photon counting along with the Poisson noise as the only source of experimental uncertainty. Then, the overcomplete frame, $\Xi^{MUB}$, has an advantage over the minimal frame, $\Xi^{SIC}$. However, the improvement in quality due to the overcomplete frame appears to be rather moderate.

Furthermore, if we increase the number of photons involved in measurements or include dark counts as another source of experimental noise, both frames deliver the same quality of pure state estimation. In such cases, the four-element minimal frame, $\Xi^{SIC}$, which consists of the vectors generating the SIC-POVM, fully suffices for QST of qubits.

In conclusion, one can agree that an overcomplete set of measurement operators is advisable only if we consider experiments with single photons and perturbation term can be neglected. This outcome is in line with other results which demonstrate that overcomplete measurements can improve the efficiency of QST frameworks \cite{Zhu2014}.

In the future, a similar approach can be applied to investigate the accuracy of different frames in higher-dimensional cases. In particular, the problem of reconstructing $4-$dimensional complex vectors shall be studied since this case includes entangled photons.

\section*{Acknowledgments}

The author acknowledges financial support from the Foundation for Polish Science (FNP) (project First Team co-financed by the European Union under the European Regional Development Fund).

\end{document}